# A Framework for Effective Corporate Communication after Cyber Security Incidents

Richard Knight [a] and Jason R. C. Nurse [b]

[a] *WMG, University of Warwick, UK*
[b] *School of Computing, University of Kent, UK*

Corresponding author: Jason R.C. Nurse, j.r.c.nurse@kent.ac.uk

**Highlights**

- We develop a framework for corporate communications in the event of a cyber incident
- Best practices for effective data breach announcement decisions identified
- The framework is grounded in a systematic review and real-world case studies
- Interviews with senior industry professionals allow framework evaluation and refinement
- The framework can complement security incident response and management in businesses

**Abstract**

A major cyber security incident can represent a cyber crisis for an organisation, in particular because of the associated risk of substantial reputational damage. As the likelihood of falling victim to a cyberattack has increased over time, so too has the need to understand exactly what is effective corporate communication after an attack, and how best to engage the concerns of customers, partners and other stakeholders. This research seeks to tackle this problem through a critical, multi-faceted investigation into the efficacy of crisis communication and public relations following a data breach. It does so by drawing on academic literature, obtained through a systematic literature review, and real-world case studies. Qualitative data analysis is used to interpret and structure the results, allowing for the development of a new, comprehensive framework for corporate communication to support companies in their preparation and response to such events. The validity of this framework is demonstrated by its evaluation through interviews with senior industry professionals, as well as a critical assessment against relevant practice and research. The framework is further refined based on these evaluations, and an updated version defined. This research represents the first grounded, comprehensive and evaluated proposal for characterising effective corporate communication after cyber security incidents.

# 1. Introduction

On the morning of 23rd October 2015, the Chief Executive of TalkTalk, a major UK telecommunications provider, featured on BBC Radio Four's Today Programme. The organisation had just discovered a data breach and subsequently wanted to inform its customers. During the radio interview, however, she had to admit to not knowing whether the data lost was encrypted (Khomami, 2015a). This apparent lack of knowledge resulted in criticism both in social (BBC Radio 4, 2015) and traditional media (Khomami, 2015b). Other public statements trying to compare the organisation's cyber security capability favourably against competitors and the application of early termination fees to those customers wishing to leave were similarly admonished (Cluley, 2015). Days later, a UK House of Commons enquiry had been convened (BBC, 2015b) and the firm was subsequently fined £400,000 by the UK Information Commissioners Office (ICO) (ICO, 2016).

Although the enquiry and ICO investigation found significant deficiencies in TalkTalk's cyber security, the organisation's approach to public communications has also drawn criticism (Maddocks, 2015) and is likely to have made the situation worse. Similar deficiencies in communication can also be witnessed more recently in other high-profile cases such as the Equifax breach in 2017, and the Travelex cyberattack in 2020. Whilst it is a key task of cyber security professionals to prevent such attacks, no system is totally secure, so it is important that if a breach occurs organisations respond appropriately. The way that businesses communicate to their customers and external stakeholders following a data breach can impact their share price and reputation to such an extent that they can be considered cyber crises (Wang and Park, 2017); these also have further implications for the business' continuity and resilience. The approaches for communication following a cyber security incident have, therefore, become an important area of research and practice.

In this paper, we seek to further academic research on the topic of appropriate corporate communication and public announcements after a cyber security incident. Through an investigation into related academic and practitioner work, we aim to improve the understanding of what constitutes effective and poor external communication following such incidents. In addition to providing this insight, a primary contribution of our research is the development, evaluation and refinement of a framework to support organisations in corporate communications after an incident. This framework can provide organisations with insight into the types of activities that are necessary after an incident (or cyber crisis), the key organisations and personnel with which to engage, and how they may consider crafting and disseminating public announcements and other correspondence. We note here that security incidents can be characterised by their malicious or unintentional nature; this study and the framework outlined consider malicious attacks and data breaches in particular given their prevalence and the resulting media interest (Berkman et al., 2018). In later sections, we also reflect on our findings and comment on the suitability of the framework to support breaches that are as a result of unintentional actions (e.g., accidental insider threats (Nurse et al., 2014)).

The approach to achieve our research aim is guided by a rigorous methodology which takes inputs from academic research and commentary on current data breach cases from cyber security specialists. It is further informed by interviews with senior, and highly experienced, security and public relations professionals.

In summary, the contributions of our work are as follows:
- A critical investigation into effective and poor communication after cyber security incidents, according to academic literature and a series of real-world case studies (including commentary from well-respected, international security specialists).
- The development, evaluation and refinement of a framework to enhance best practice regarding corporate communications and announcements in such situations.

The remainder of this article is structured as follows. In Section 2 we scope our research and review the background theory on data breach communications. The research methodology is outlined in Section 3. We present the first stage of our work, a systematic, and critical review of current literature on the topic of data breach communications in Section 4. This is followed by an analysis of real-world cyber incident communications (and related commentary), drawn from several case studies in Section 5. Section 6 introduces our proposed framework to enhance incident response communication strategies in organisations, which is then evaluated and appropriately updated in Section 7. Finally, we conclude and present avenues for future work in Section 8.

# 2. Background and Related Work

## 2.1 Cyber Security Incidents and Communications Theory

Security incidents can take various forms. They can encompass the accidental exposure or loss of data, ransomware attacks, or disruption of systems (Sarabi et al., 2016). Although data breaches can be seen as a subset, there is variation over their definition (Schatz and Bashroush, 2016; Edwards et al., 2016). For our research, we focus the term data breach on unlawful acts and use an amended version of the ISO description accordingly: *"compromise of security that leads to the unlawful destruction, loss, alteration, unauthorized disclosure of, or access to, protected data transmitted, stored or otherwise processed"* (Adapted from: ISO, 2015). As mentioned earlier, we scope our work at this point to malicious data breaches.

Another fundamental term in our research is that of a crisis, or cyber crisis. The loss of a significant amount of personal or sensitive data can be extremely detrimental for organisations given its impact (Agrafiotis et al., 2018; Wang and Park, 2017). Whilst various definitions of a crisis have been put forward (Mitroff, 1988), the following appears to encapsulate their essence: *"A high consequence, low probability [event], overlaid with risk and uncertainty, conducted under time-pressure, disruptive of normal business and potentially lethally damaging to organizational reputation"* (Gregory, 2005, p. 313). Although this description was developed through work in another domain, it can aptly be applied to describe the aftermath of a substantial data breach (which can also be linked to the size of the breach or the type of data lost). A cyber crisis can, therefore, be seen as a crisis resulting from a data breach or similar security event.

The field of public relations has considerable research on crisis communications; situational crisis communication theory (SCCT) (Coombs, 2007) and theory of image restoration (Benoit, 1997) being the preeminent models in this area (Avery et al., 2010). They provide a framework for the types of response available to organisations during a crisis. Whilst these concepts are well regarded, they have been criticised for not being practical, meaning they allow practitioners to understand the approach they are using but do not provide detailed criteria that would enable organisations to determine which approach to use (Avery et al., 2010). This is particularly true where data breaches are concerned, due to the ambiguity over whether the organisation is a victim as a result of being hacked or is responsible as a result of having inadequate security measures in place (Bentley et al., 2018).

## 2.2 External Communication after a Data Breach

Research into how businesses communicate during a cyber crisis has tended to take a top-down approach. Kim et al. (2017) for example, apply a deductive method to investigate how newspapers interpret data breach corporate communication. This has the potential to constrain thinking, as analysis of crisis communication is framed by these models. For instance, some studies into data breach communication models have predetermined SCCT as the framework applied to case studies (Wang and Park, 2017; Wang and Johnson, 2018), which means that the data are examined through a fixed set of categories which may overlook other explanations for the results. Qualitative methods such as qualitative thematic analysis provide an alternative approach which allows patterns to be observed that may be obscured by a more constricted method (Kaefer et al., 2015; Strauss and Corbin, 1994).

Much of the research carried out on data breaches is focused on companies based in the USA (Spanos and Angelis, 2016). In particular, the primary studies on data breach communications found during an initial literature review have used USA corporations as case studies (Wang and Johnson, 2018; Wang and Park, 2017). There is also notably work exploring the relationships between social media, stock prices and data breach announcements on USA companies (Rosati et al., 2019). Although the USA has a substantial corporate base, this focus illustrates a gap in research. We posit that other countries, such as the UK for instance, could be another intriguing case for research given its position as a major world economy with extensive internet use amongst its population. In this paper therefore, we seek to provide another perspective on the problem of corporate communications after a data breach. In subsequent sections we provide a more critical review of existing literature and detail how we analyse it, and our own primary data to develop the new framework.

# 3. Methodology

The research presented in this article investigates a topical problem that has been explored and discussed in research and practice. As such, our methodology is grounded in insights from both of these domains, and based on those foundations and contributions, we arrive at our findings and framework. The methodology has four main steps, which are described below and depicted in Figure 1.

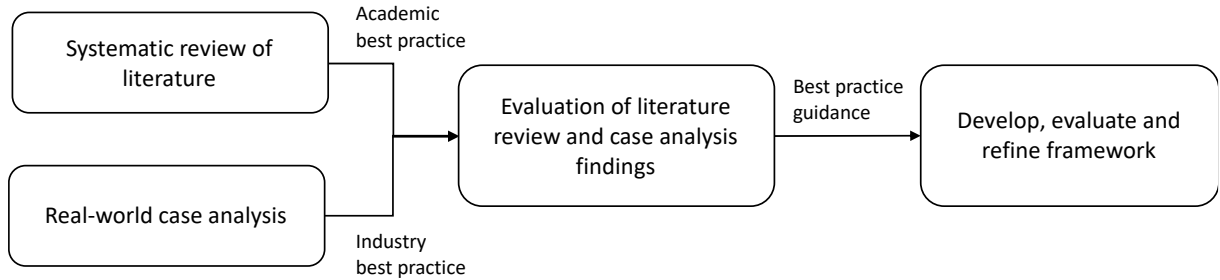

Figure 1 – Research methodology overview

## 3.1 Systematic Review of Literature

Existing academic literature is of great value in informing what response and activities companies should adopt after a data breach. As such, we first critically review literature in the field of crisis communication to understand best, and recommended, practices. For this review, we use the PRISMA framework due to its ability to provide rigor, consistency and transparency in the selection of relevant articles (Moher et al., 2009); PRISMA has also been used on several occasions in computer security research (Fernández-Alemán et al., 2013).

In PRISMA, decisions are required on article sampling and filtering criteria. Considering that our research spans across the computer security and business fields, we scope the databases and articles sampled to the sciences and business (see Section 4.1 for the list). Similarly, as our scope is on corporate communication after a data breach, articles should be relevant to this intersection of topics. Once selected, each article is examined using thematic analysis (Braun and Clarke, 2006), and a series of codes and themes produced which provide insight into the article and the relevance of its findings. Further detail is provided in Section 4.

## 3.2 Real-world Case Analysis

To complement the literature review, we next investigate cases of real-world data breaches to understand company responses in the context of commentary from industry specialists; in this commentary, we particularly consider the efficacy of the communication that occurred. While there are various sources which catalogue data breach cases (e.g., Morgan, 2018), official sources are preferable given their reputation and reliability. To assist in the sampling of cases, we scope our research to the UK. This decision is based on the reality that most research (e.g., Wang and Johnson, 2018; Wang and Park, 2017) has focused on the USA, and that the UK is another major economy (in Europe and worldwide) worthy of study in general and for comparison. Based on these criteria, the UK's National Cyber Security Centre (NCSC) and, in particular, its Threat Report publication (NCSC, 2019a) is selected as a source for cases. These reports are published regularly and contain information on major cyber security breaches, including incidents that may not necessarily be subject to investigation. While the NCSC does not publish the specific criteria used to select incidents, thus making it challenging to comment on completeness, given its remit (to provide guidance on UK cyber security and its insights into cyberattacks targeting businesses across the country) we view it as a reliable and robust source of pertinent incidents.

Having established the sampling approach for the data breaches, a similar mechanism is required for ascertaining security commentators who reflect on breaches and the response of the organisation. A robust method is essential to identify a suitable set of sources to ensure the credibility of the assessment of what may be regarded as effective and poor communication. While we could openly search online for commentators, we note that references to security commentators are also often included in the NCSC threat reports. Although these tend to reference a single source, it is possible, by analysing all of the commentators cited, to obtain a set of commentators who can be used as sources for security commentary on the selected cases. The NCSC threat reports are therefore selected as a source. Similar to the literature review, thematic analysis is applied to data gathered in this second stage to provide further insight pertinent to our research aim. We provide additional detail in Section 5.

### 3.3 Evaluation of Literature Review and Case Analyses

Based on the findings from academic literature (including suggested best practice) and real-life scenarios (i.e., from security commentators) we look to develop an improved understanding of what factors may drive effective and poor crisis communication following a malicious cyber security breach. The themes and observations gathered from the systematic literature review and case analyses are therefore compared and evaluated in order to establish consistencies and differences. This also enables an assessment of how current practice found in the case studies differs from, or follows, the best practice in the literature. We apply the technique of cognitive mapping (Miles et al., 2014) to allow the themes, patterns and their relationships to be examined and more clearly compared. This, therefore, provides the foundation for our key findings and the framework's development. We expand on this work in Section 6.

### 3.4 Develop, Evaluate and Refine Framework

Using the insights from earlier research stages, we create a framework that is able to inform organisations in the preparation and execution of cyber incident response plans pertaining to corporate communications. The framework combines best practice and recommendations in a structured, coherent format suitable for businesses in planning for and reacting to a cyber security breach. To evaluate our proposals, we conduct a series of semi-structured interviews with senior industry practitioners (e.g., potential users of the framework), including CISO-level, and also experts involved with crisis communication after a cyber security incident (e.g., directors and leads in crisis management organisations); this latter group is perfectly placed to comment on the framework's contributions. Interviews are used in preference to other methods, such as questionnaires, as they allow the capture of in-depth feedback (Burdon and Coles-Kemp, 2019). Interview findings are then analysed using thematic analysis and the framework is refined to incorporate these insights. Additional information is provided in Sections 6 and 7.

# 4. Reviewing Cyber Crisis Communication Literature

## 4.1 Sampling Strategy

Academic literature can provide a strong foundation for an investigation. In the case of communication and public relations following a data breach, however, an initial literature review found a limited body of work on the subject. Indeed, this is underscored by Gwebu et al. (2018) who highlight the lack of understanding of how organisations should respond effectively to a data breach. In our systematic review therefore, we sought to adopt a broad and more inclusive method to increase the likelihood of finding suitable articles.

In the context of the PRISMA approach (Moher et al., 2009), we first identified a series of databases to search and search terms to use. The databases selected were Business Source Complete, Web of Science, Science Direct, ACM Digital Library, Emerald Insight, and IEEE Xplore. These are well-recognised databases and have been used in similar systematic literature reviews in the field of cyber security (Spanos and Angelis, 2016; Lezzi et al., 2018). They also allow the identification of peer-reviewed articles, which increases the likelihood of finding high quality, objective contributions. As our research is focused broadly on security, incidents and communication, these formed the core topics of the search terms, and synonyms and alternative terms were used for completeness. The final search string used was: *("cyber security" OR "cybersecurity" OR "systems security" OR "network security" OR "information security" OR "cyber crisis") AND ("data breach" OR "hack" OR "incident") AND ("communication" OR "announcement" OR "stakeholder management" OR "notification")*. This query was executed in each database, and where possible, applied to the title, abstract or keywords; this is a common technique (Fernández-Alemán et al., 2013; Lezzi et al., 2018) and increases the likelihood of finding relevant articles.

In total, 3516 articles were found; following deduplication, this was reduced to 3405 articles. In line with PRISMA, the dataset was screened to ensure only relevant articles were considered. This consisted of a pre-screening activity, where article titles, abstracts and full-texts were reviewed for relevance. This resulted in 197 articles. Two eligibility criteria were then applied. Firstly, only articles from well-regarded and cited journals were included. This was to increase the likelihood that the credibility of the literature would be high. The Journal Citation Report (JCR) (Clarivate Analytics, 2019) provides a mechanism to compare journals based on citation data and although this does not in itself determine the value of a publication, it establishes that other authors are referencing its papers. In order, therefore, to ensure that academically robust articles were used in this study, those without JCR values were discarded. Secondly, articles published before January 2009 were excluded. Although the quality of an article cannot be seen as being related to its age, changes in practice, technology and legislation can make its findings immaterial, hence our decision. These eligibility criteria excluded a further 152 articles resulting in a final set of 45.

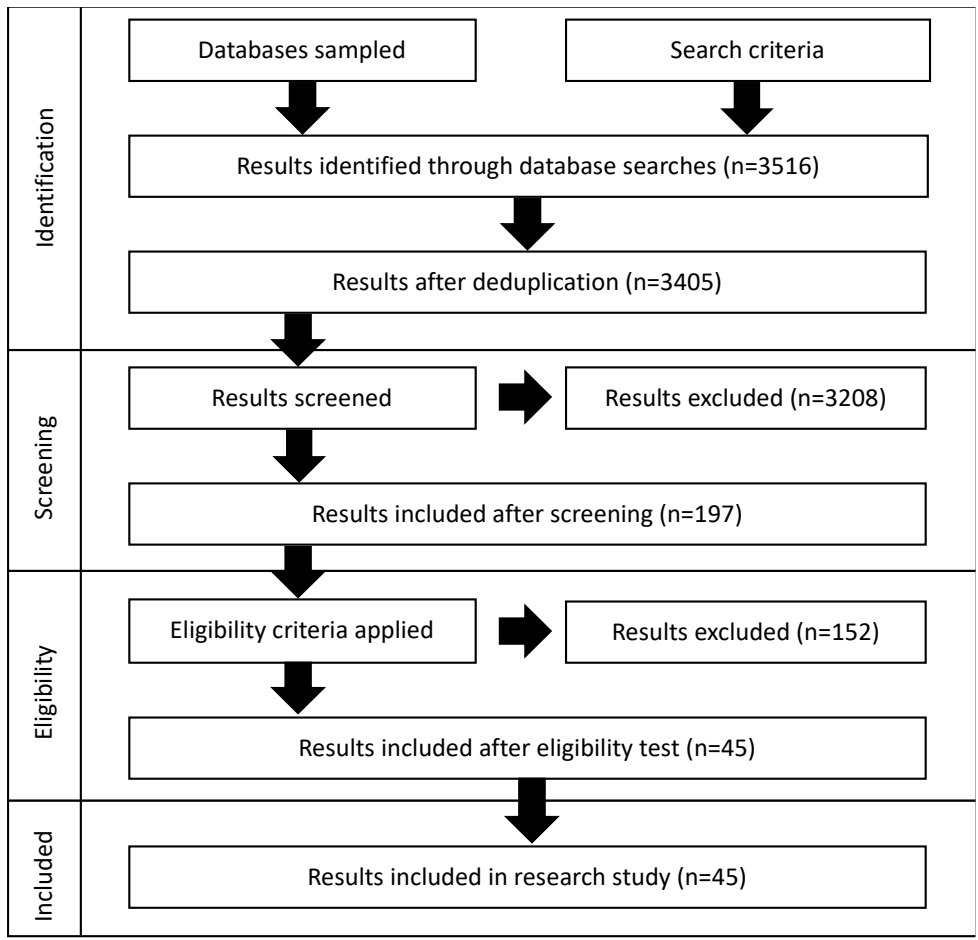

Figure 2 – PRISMA flow diagram

In Figure 2, the sampling strategy, including totals of articles excluded at each stage are summarised.

## 4.2 Extracting Research Themes

The articles identified were then analysed using the thematic analysis approach outlined by Braun and Clarke (2006). The process of familiarisation and code generation was performed against the articles with the resulting codes grouped into an initial set of categories. Although the research aim provided the context for the analysis, it was not used to restrict the code generation process. Areas of research that may help inform the debate around effective communication following a data breach are potentially multidisciplinary, therefore any attempt to restrict coding at an early stage could overlook noteworthy insights. The code generation approach consequently resulted in a broad range of categories representing several domains. These were subsequently reviewed to identify themes. To explore these further, cognitive mapping was used as it allows data to be visualised and commonalities and patterns more clearly identified (Miles et al., 2014).

## 4.3 Analysis of Key Findings from Cyber Crisis Communication Literature

A sample of the themes identified is presented in Table 1, along with the number of articles and codes related to each theme. This provides a useful indicator of how many times codes connected with each theme were identified and the breadth of articles they were found in.

| Theme | Articles (Codes) |
|---|---|
| Stock market reaction | 15 (54) |
| Legal Requirement to Notify | 11 (47) |
| Message Framing | 12 (50) |
| Not Disclosing | 11 (17) |
| Protecting Reputation | 11 (36) |
| Negative Emotions | 6 (24) |
| Word of Mouth | 3 (7) |
| Complexities with Outsourced Functions | 1 (6) |

Table 1 – Sample of themes derived from the review

Below we discuss a selection of the most pertinent of these themes for our research, followed by a more general reflection on other relevant themes.

*4.3.1 Stock Market Reaction*

A significant number of articles focused on the stock market impact of data breaches. Most, but not all articles reported an initial drop in share price following a data breach. The temporary nature of this drop was seen by some researchers as the result of the "knee jerk" reaction of ill-informed investors, rather than a product of the organisation's effective or poor communications approach (Wang et al., 2013). Other factors such as the size of the data breach, the company's industry segment (Hinz et al., 2015) and the nature of its products (Jeong et al., 2018) were also highlighted. One article, however, argued that stock market reaction was greater for organisations that already had poor reputations. It further asserted that their use of communications strategies to demonstrate that the organisation is committed to stakeholders and to addressing the problem can reduce the impact on its share price (Gwebu et al., 2018). Whilst this is only a single article, it does suggest that 'image renewal' strategies such as these may be effective and beneficial to companies, particularly those already perceived as tainted by the public.

*4.3.2 Legal Requirement to Notify*

The regulatory environment associated with data breaches was discussed by several researchers. In particular, the requirement to notify data protection authorities and impacted data subjects were assessed. The General Data Protection Regulation (GDPR), for instance, stipulates communication should be used to mitigate the risk of harm to individuals but that this should be balanced against ensuring the cost of this to the organisation is not too overwhelming (de Hert and Papakonstantinou, 2016). Other studies highlighted that adequate encryption of the data lost may be sufficient, thus suggesting that communication to data subjects is not always required (Nieuwesteeg and Faure, 2018). Also of interest were studies that considered other jurisdictions, most notably Australia and the United States (Burdon et al., 2010; Rosati et al., 2017). These, combined with other articles that draw attention to legislation for specific industry segments such as telecommunications, suggest that whilst GDPR provides a legal basis for disclosure of a data breach to the public, other legislation may apply, particularly where people in other countries are impacted. This multi-jurisdictional complexity regarding disclosure rules should therefore be considered as part of an effective communications approach.

*4.3.3 Message Framing*

Of particular interest are the insights provided on the framing of communications following a data breach. Messaging framing can be seen as a method for distilling complex ideas into clear, compelling and digestible dialogue (de Bruijn and Janssen, 2017). This can be used following a breach to ensure stakeholders are informed appropriately. Whilst the use of such framing strategies is central to the SCCT approach (Syed, 2018), other techniques such as Error Management (Zhang et al., 2019) and Extended Parallel Processing Model (EPPM) (Oh et al., 2018) were also identified as potential framing strategies. These alternatives, therefore, may need to be reviewed alongside SCCT as organisations decide what is a suitable approach to communication.

*4.3.4 Other Themes*

The emotional impact of the data breach on those affected was discussed by multiple researchers (Chen and Jai, 2018; Zhang et al., 2019; Syed, 2018; Janakiraman et al., 2018). The vulnerability felt by individuals impacted by a breach was highlighted (Janakiraman et al., 2018), as were the different negative emotions directed at organisations dependent on the crisis stage and where the perceived blame was attributed (Syed, 2018). A link was also determined between the resulting negative emotions and 'negative word of mouth' which could in turn damage the organisation's reputation. It can, therefore, be seen as crucial that an effective communications strategy recognise and attempt to alleviate the negative feelings of those affected.

Another point that warrants discussion is the complexities associated with outsourced functions. This was investigated by Modi et al. (2015) who highlight the difficulties in coordinating responses and the potential negative public reaction due to customers often being unaware of the third parties involved in their transactions. This can often lead to strained relations with suppliers and to a greater impact on the company's share price. This view is complemented by Porcedda (2018) who describes the care that should be taken by organisations processing sensitive data, such as hospitals when dealing with breaches involving cloud-service providers. Although not seen as a complete solution, the use of contracts to establish responsibilities was put forward as beneficial. This may have a significant bearing on data breach communications and could, therefore, be considered as part of an effective approach.

## 5. Analysing Real-world Data Breach Communications

### 5.1 Sampling Strategy

As introduced in Section 3, NCSC threat reports were identified as an appropriate source of both notable data breaches within UK companies, and for the definition of relevant cyber security commentators. Our sampling strategy therefore began by collecting the incidents identified by these reports; this resulted in a listing of 128 potential cases from threat reports between 23 September 2016 and 3 May 2019 inclusive. Cases were then screened to establish whether they were associated with data breaches and were UK-centric. We interpreted UK-centric to mean that the organisation could be considered as a UK company or that UK individuals were primarily impacted by the breach. Therefore, organisations such as British Airways (BA), which is regarded as the 'British Flag Carrier', would be included even though they are ultimately owned by Spain's IAG Group. Table 2 lists the data breaches defined.

| Organisation | Month Breach Reported |
| --- | --- |
| Deliveroo | January 2019 |
| B&Q | January 2019 |
| British Airways (BA) | September 2018 |
| Superdrug | August 2018 |
| Butlins | August 2018 |
| Dixons Carphone | June 2018 |
| Clarksons | November 2017 |
| PageUp (Whitbread) | June 2018 |
| Ticketmaster | June 2018 |
| Great Western Rail | April 2018 |
| Sodexo | February 2018 |
| National Lottery - Camelot | March 2018 |
| Deloitte | September 2017 |
| Hotpoint UK | April 2017 |

Table 2 – UK-Centric data breaches identified

To identify suitable commentators, a similar approach to the method for selecting cases was used. This involved reviewing the threat reports sampled for references that represent potential third-party sources. These could be explicit, such as "*Security researcher Chris Vickery has reported…*" (NCSC, 2017), or a reference via a hyperlink. Through this review, we identified 229 references to third parties. Each of these was then reviewed and categorised according to whether it was commentary from a security specialist, a report from a government body, news agency or security company, or a self-report from the organisation breached. Where possible commentators were identified, an assessment of their website and other online reports was conducted to establish whether they could be considered as appropriate; for instance, if they previously covered security issues, possess a reputation in the security field, etc. The commentators selected were: Schneier on Security, Brian Krebs, Troy Hunt, Chris Vickery, Oliver Hough, Gabor Szathmari, We Live Security, Cyberscoop and Ars Technica. Having determined both the case studies and the commentators, a search of each of the commentator's websites was performed. This allowed us to gather any relevant incident analysis or discussion on each case from all of these sources.

## 5.2 Analysis of Key Findings from Commentator Reviews

The analysis of commentator data was conducted using the same process as outlined in Section 3.2, and led to several important themes being defined. We present a sample of the themes in Table 3, along with the number of commentaries and codes related to each theme.

| Theme | Commentaries (Codes) |
| --- | --- |
| Credibility of Statement | 12 (22) |
| Downplaying | 6 (8) |
| Media Spin | 6 (9) |
| Not Disclosing or Delaying | 4 (9) |
| Focus on the Customer | 6 (6) |
| Previous Breaches in Same Company | 3 (3) |
| Admission of responsibility | 2 (2) |

Table 3 – Sample of themes derived from commentator reviews

In what follows, we briefly explore some of the most relevant themes to our research.

*5.2.1 Credibility of Statement*

The credibility of statements made by organisations suffering a data breach was questioned by commentators. Hunt, for instance, repeatedly stated "*I find this a little bit hard to believe*" when discussing the statement made by the UK National Lottery recommending their 10.5 million users reset their user IDs and passwords following reported suspicious activity on 150 accounts (Hunt, 2018b). BA's account of their own data breach timeline was also questioned by researchers who were able to analyse the certificate used by hackers to mimic its website (O'Donnell, 2018b). Finally, the actions of Ticketmaster in attempting to blame a third-party supplier resulted in a strong public response from them (Shoorbajee, 2018b). In that case, after being accused of providing insecure software, the third-party publicly responded by arguing that Ticketmaster had failed to notify them that Ticketmaster were deploying it to their payments page; and that, if they had known, they would have advised it was insecure. They also added that Ticketmaster was the only customer impacted. Validating statements to ensure they make sense and are backed up in fact, can, therefore, be seen as an important element of effective data-breach communication. It can also be argued that blaming partners may incite public disagreements which may be considered an ill-advised crisis strategy.

*5.2.2 Not Disclosing or Delaying*

The timing of communication relative to the data breach and whether the organisation chooses to inform the public may also impact credibility. In the cases of Dixons Carphone and BA, questions were raised over whether the public was informed in a timely manner. Deloitte, however, appeared only to find out they had been breached months after the event. Reports by Cyberscoop (Bing, 2017) also suggest that the story was first made public by The Guardian news outlet rather than the company itself. This reluctance to admit they had been compromised is understandable given Deloitte provides cyber security services to many major organisations (Krebs, 2017). That being said, the delay may have prompted the whistle-blower that informed The Guardian to act, resulting in negative publicity for the firm.

From the case of Deloitte, it may be argued that organisations should aim to communicate soon after they discover the data breach. This would however suggest that even if only partial information is available, the firm may have to estimate the number of impacted parties. This seems to have been the case for BA who had to revise their numbers upwards from 380,000 (Foltyn, 2018a) to 429,000 (Seals, 2018a). This topic highlights the dynamic between the need to communicate quickly and the time required to understand the situation and provide accurate information. It is thus important that the timing of external communication is considered as part of any effective guidance.

*5.2.3 Focus on the Customer*

Commentators also discussed the steps firms were taking to protect those impacted by the breach. For instance, a Threatpost commentator summarised the steps BA was taking to ensure customers were not out of pocket (Foltyn, 2018a). This type of approach involving communicating compensation facilities to customers if they are subject to fraud was, however, criticised by another commentator for 'causing complacency' (Anscombe, 2019). He describes his own experience where his credit card was used for money laundering because, he implies, his details had been stolen months earlier as part of the BA breach. Having been subject to this type of crime, his view was that organisations need to communicate more about what they are doing to pursue the criminals. This theme of 'facing up to the bad guys' was reinforced by another commentator's praise for Clarkson's response to its breach by informing the police and accelerating its implementation of additional security measures (Cluley, 2017). Consequently, it may be concluded that such offers of compensation and protection should be outlined in communications along with a determination to support law enforcement in order to track down the perpetrators.

*5.2.4 Other Themes*

Downplaying was a tactic used by a number of companies in an attempt to reduce the perceived significance of the attack. Deloitte, in particular, was reported as trying to 'downplay' their data breach by both Krebs (2017) and Schneier (2017). BA, meanwhile, highlighted that there was no evidence of fraud (Shoorbajee, 2018a), whilst Dixons Carphone stated that the data did not include PIN or CVV information (Foltyn, 2018b). Whilst this may be a statement of fact at that time, this position is undermined by the personal stories of loss after the event, such as that by Anscombe as discussed previously, and by Hunt (2018a), whose Have I Been Pwned site (https://haveibeenpwned.com) demonstrates that stolen data is available to be used long after the breach itself. As this is likely to be recognised by media and the general public alike, such an approach may be counter-productive.

Many of the organisations studied apologised for the data breach. Clarkson's CEO "sincerely apologise[d] for any concern" due to the incident (Cluley, 2017) and the CEO of Dixons Carphone admitted the company had 'fallen short' (O'Donnell, 2018a). Admission of responsibility and work to address security weaknesses was seen as key to restoring consumer trust by a report cited by Threatpost (Seals, 2018b). This suggests that this type of response, whilst uncomfortable for organisations, may have some merit in the long term.

Multiple commentators reflected on previous breaches that the impacted companies, notably BA (Foltyn, 2018a) and Dixons Carphone (Foltyn, 2018b), had suffered; this was possibly trying to imply they had not learnt lessons from prior incidents. This was particularly salient in cases where current post-breach communications had claimed to be sorry for the new breach and leak of customer data. From an effective communications perspective, a noteworthy point is that organisations need to be careful in how messages are crafted especially in situations where they have been subject to incidents in the past. Next, we use the findings of this analysis of commentaries and the systematic review to develop the framework.

# 6. Developing a Framework for Effective Corporate Communications

From the analyses in Sections 4 and 5, several key points have emerged which can be related to effective and poor corporate communication following a data breach. These are further evaluated and compared in this section in order to understand whether there is a consistency between the guidance derived from these sources. Using these findings, a framework is proposed that seeks to provide support to organisations on how they can provide effective external communication following a malicious data breach.

## 6.1 Analysis and Comparison of Findings

In order to assess the findings thus far, a cognitive mapping method was used (Miles et al., 2014). This provided a clear representation of the abstracted data from the two prior analyses, allowing the visual examination of results. Several themes identified were considered particularly relevant to our research aim, and were further enhanced to develop our framework. Below, we present a selection of the some of these to provide insights into how our framework was built.

*6.1.1 Blame and Situational Crisis Communication Theory (SCCT)*

SCCT provides a framework of categories of crises which allows organisations to select a communication response strategy appropriate to a given scenario (Coombs and Holladay, 2002). In terms of data breaches, an experiment by Chen and Jai (2018) determined a partial correlation between reduced trust and organisational responsibility for a breach. This is supplemented by Syed (2018) who posits that perceived accountability for the loss of personal information can lead to negative emotions potentially resulting in deprecatory social media posts from impacted individuals. Both pieces of research, however, show variability in participant responses, with Syed identifying changes in emotional response depending on the crisis stage. It is also important to recognise the role that the media has in framing public perception. Chen and Jai (2018) for instance, argued that by receiving a data breach message via the media, individuals are more likely to consider the enterprise culpable. This may be seen as significant, given the confidence people can have in these sources (Kulikova et al., 2012).

When applied to data breaches, the approach advocated by SCCT of establishing responsibility may lead organisations to look at the perpetrators or other parties as culpable and see themselves as victims. This appears to be the case in the TalkTalk example where the CEO highlighted that she had been a victim herself (BBC, 2015a). Though within this study, this technique was only used by Ticketmaster who attempted to blame a supplier, this hypothesis is supported by other research which has established this as a method used by businesses (Wang and Johnson, 2018; Wang and Park, 2017). It can, therefore, be concluded that it is still utilised for such incidents.

The results of apportioning blame can be seen from the commentary associated with the Ticketmaster breach, where a conflicting and seemingly plausible account was given by the supplier (Shoorbajee, 2018b). Such public disagreements can only be detrimental to a business' attempt to protect its reputation. From the commentator analysis, the use of apologies and actions to mitigate the risk of harm to the data subject have been the subject of praise. This suggests that organisations that take responsibility and are seen to be proactively trying to address the problem are perceived in a positive light. This is supported by some of the research found in the systematic literature review (Janakiraman et al., 2018; Soomro et al., 2019). In particular, Wei et al. (2017) found that error management techniques could be used following a data breach to improve customer perceptions and this appears to be corroborated by further studies (Zhang et al., 2019; Berezina et al., 2012).

Although the value of taking responsibility appears clear, a number of themes were identified in both data sets with regards to the risk of litigation. This is brought into context by the in excess of 240 lawsuits raised against Equifax following their data breach in 2017 (Berkman et al., 2018). Dependent on the jurisdiction, these actions can result in a significant cost to an organisation in terms of damages and legal fees. This has, however, to be balanced against the onerous sanctions that can be applied against businesses should authorities find they have not followed their obligations to inform impacted data subjects (ICO, 2019a). It is also important to recognise that class actions do not necessarily fall on the side of the plaintiff, with the burden of proof in some regions, notably the USA, requiring evidence of actual harm as a result of the breach. This can be difficult to ascertain, given the amount of personal data freely available due to other data breaches. It can thus be argued, that the risk of litigation is now outweighed by the positive effects of organisations taking responsibility for the crisis and the reduced risk of fines from relevant Data Protection Authorities (DPAs).

*6.1.2 Security Basics*

The protection of data subjects by organisations has been seen as an important consideration following a breach. The commentary from Hunt around the intentions of the US Federal Trade Commission (FTC) with regards culpability for credential stuffing attacks, however, suggests that companies should be mindful of key security controls in their data breach crisis planning (Hunt, 2018c). This is reinforced by the disclosure rules, which show that within some jurisdictions the use of up-to-date encryption can mitigate the risk of harm sufficiently so that firms do not need to notify impacted individuals. Without such mechanisms in place, corporations may be subject to criticism, particularly if they have not learnt lessons from similar previous attacks (Syed, 2018; Hawkins, 2015). Our analysis in Section 5 determined that the media is very likely to emphasize previous failures with a resulting impact on reputations. As part of their crisis communication planning therefore, businesses should ensure the security basics against which they may be held to account are in place.

*6.1.3 External Partners*

The public disagreement between Ticketmaster and its supplier, discussed earlier, highlights the multifaceted arrangements that are often in place for modern products and services. This can make dealing with the aftermath of a data breach more complex and in particular, requires communication to be coordinated between parties. This can be difficult given the potential for strained relationships. Where practicable, it is therefore advisable to ensure contracts with service partners include provisions for data breaches and that they are involved with data breach crisis planning.

*6.1.4 Borderless Nature of the Internet*

An article identified as part of the commentator analysis claimed that BA was faced with a potential $650 million class-action lawsuit (O'Donnell, 2018b). Although not explicitly stated, this suggests that customers from the USA were impacted. This reinforces the earlier point— the internet is borderless and the data subjects impacted by a breach may reside in or be protected by the laws of multiple jurisdictions. This can have a number of implications for post-data breach crisis communications, in particular, whether data subjects need to be notified and the timing of such disclosure. Given the complexity of jurisdictions within some regions, it is therefore important that organisations understand in advance their legal obligations in the countries in which they trade.

## 6.2 Development of the Framework

From a detailed analysis and comparison of the key points pertaining to effective communication, we developed a framework aimed at supporting organisations in their approach towards external communication after a data breach. The process of cognitive mapping allowed related themes to be grouped together providing further abstractions. For instance, multiple themes from the systematic review were found to be associated with possible 'organisation aims', and therefore were grouped. Through a review of the associated codes and source literature, these were found to be possible organisational aims post-breach and, because this was supported by evidence from the case studies, they were included in the framework under the title 'Establish aims post breach'.

In cases where contradictory effective communications guidance was evident, an assessment was made of the weight of the opposing viewpoints in terms of the supporting academic articles, whether the theme was present in both results and, if necessary, against further scholarly or industry sources. The guidance of 'accept responsibility' (after a breach) is such an example. The concept of organisational responsibility for personal data was found in themes within both result sets. Whilst some contradictory data was found in the 'risk of being sued' theme, this was negated by codes in the same theme suggesting that the risk of litigation is reduced if data breaches are disclosed early and that there was a burden of proof for plaintiffs. This weight of evidence meant this theme was included in the framework within a section titled, "Frame the Message". We adopted similar techniques to construct the broader framework.

We present the framework that resulted from our analysis in Figures 3, 4 and 5; three figures were initially used for ease of presentation. The framework is split into two main areas, and each has several best practice guidance points as well as aspects that organisations need to consider. Here, we remind readers that this is an initial version of the framework which will be assessed and refined later.

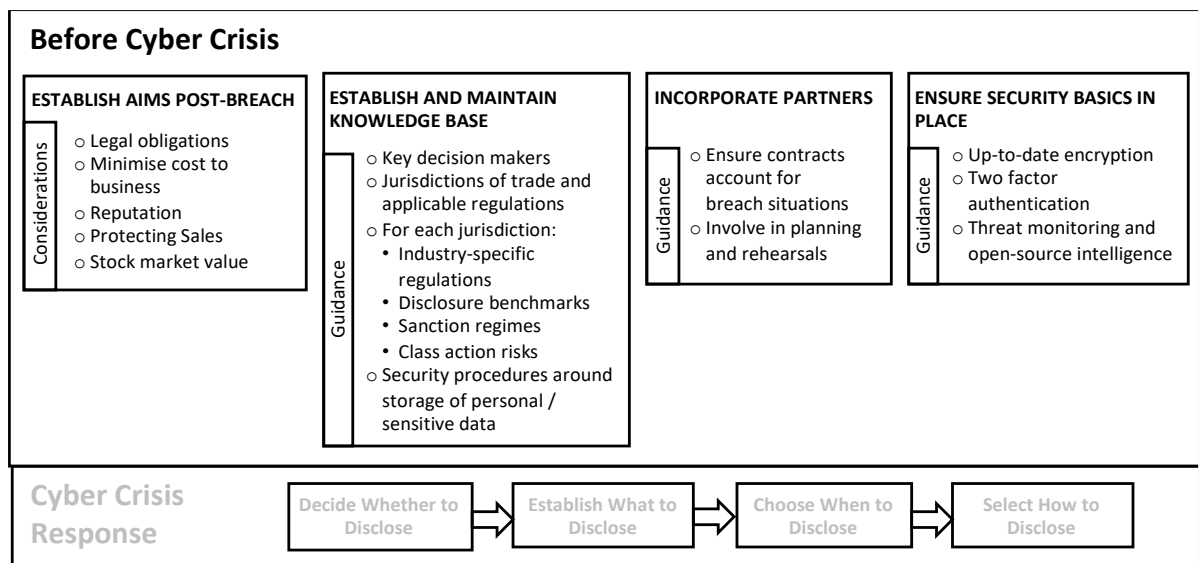

Figure 3 – Framework: Before Cyber Crisis

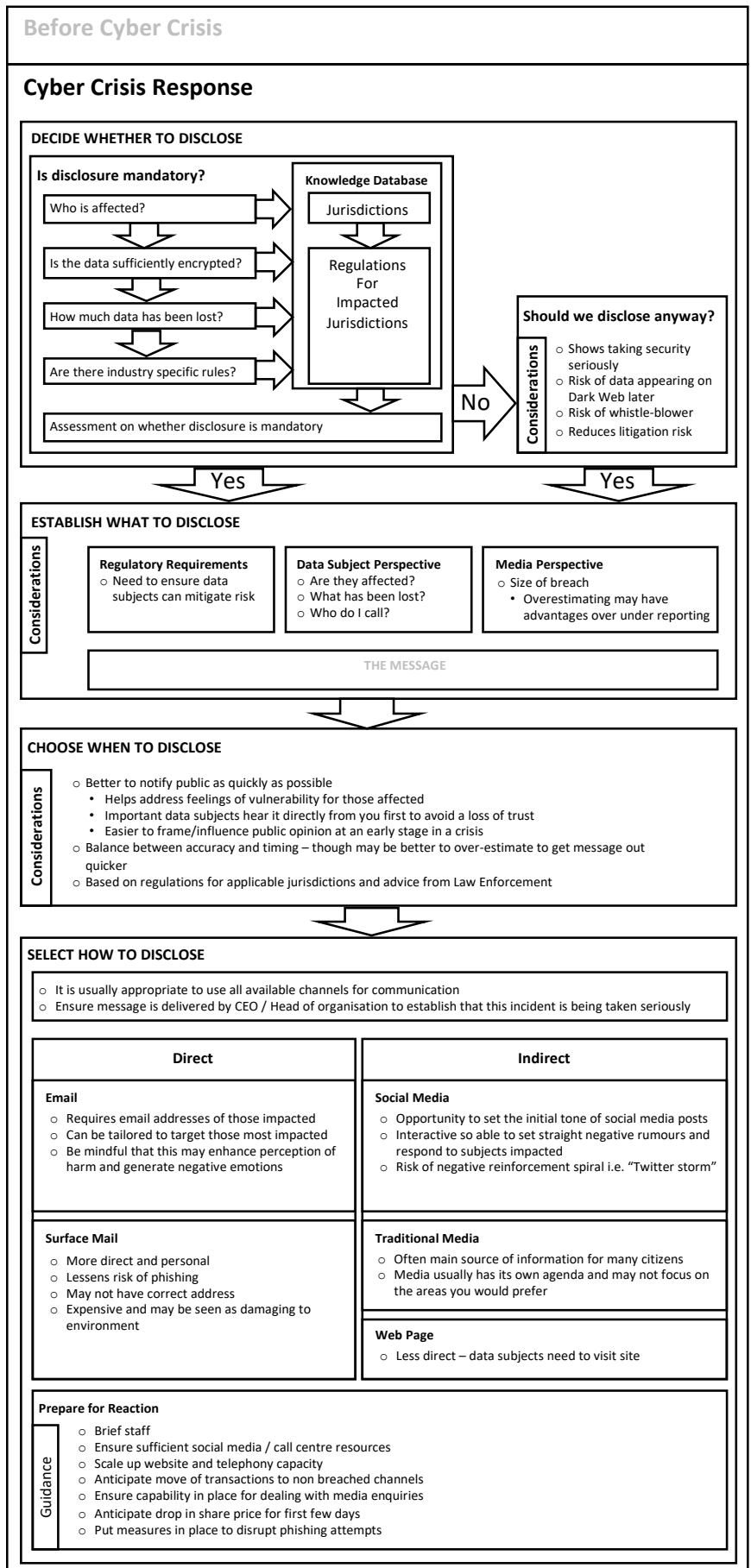

Figure 4 – Framework: Cyber Crisis Response (focusing on decision contexts)

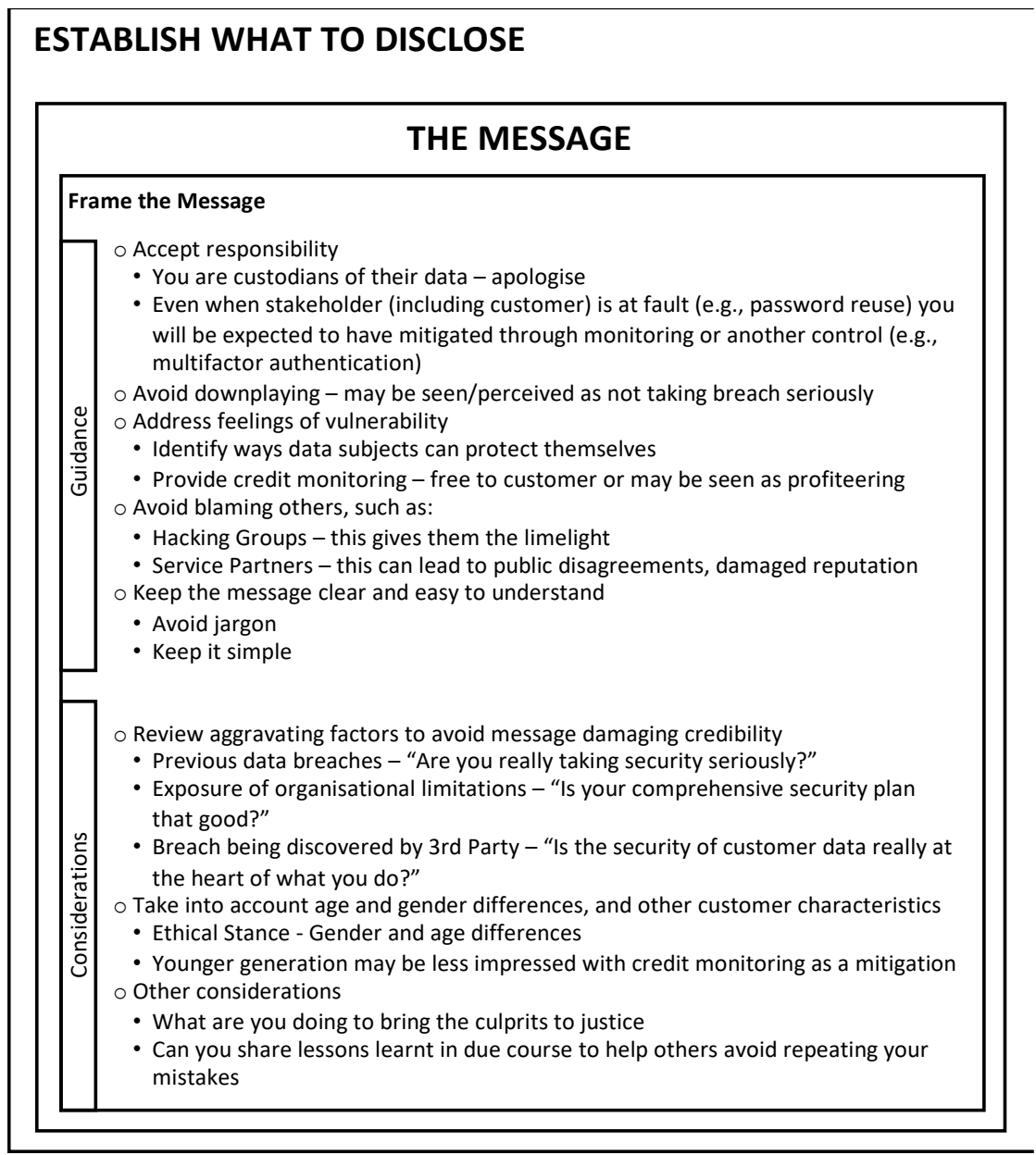

Figure 5 – Framework: Cyber Crisis Response (focusing on how to frame the data breach message)

The first area, Before Cyber Crisis (Figure 3), focuses on planning activities that must be engaged in preparation for the possibility of a data breach. It includes organisations establishing post-breach aims, identifying an updated knowledge base and incorporating business partners in plans. The second area, Cyber Crisis Response (Figures 4 and 5), concentrates on the activities necessary in the case of a data breach crisis event (with the assumption that sensitive or personal data has been breached). A disclosure decision model is outlined which presents a series of steps to assist organisations in deciding whether, what (e.g., framing the communications), when and how to disclose an incident. Together, the various points presented in these figures seek to distil the best practices on effective corporate communication surrounding a data breach.

Briefly reflecting on the framework, many of its components cover general corporate communications and announcements, such as disclosure means, timelines and message content. These are areas that are likely to also be crucial in incidents where businesses need to communicate or announce an unintentional data breach, i.e., one without a malicious nature. The most significant difference may be in the discussion about blame (see 'frame the message' in Figure 3). With incidents involving an unintentional exposure of data, typically the organisation (via its employees or stakeholders) is indisputably at fault and thus cannot reassign blame away from itself or act as a victim. In such cases therefore, it may be even more crucial to quickly assume responsibility and act. Consequently, while it is not possible to comment definitively on the applicability of the framework in supporting unintentional data breaches given our scope, it may be promising.

# 7. Validating and Refining the Framework

## 7.1 Participants

Whilst based on robust methods, the aforementioned framework and findings it embodies are theoretical in nature. In order to evaluate whether it addressed the gap identified in industry practice, we therefore sought to validate it through semi-structured interviews with senior industry professionals, both in the cyber security and public relations and communications fields. Participants were selected using a purposive sampling approach (Etikan et al., 2016) due to the impracticability of obtaining relevant individuals of the necessary seniority and expertise using probability sampling methods. This method can be seen as an effective way to obtain 'hard to reach populations' and has been used in similar studies (Unkelos-Shpigel et al., 2015).

In total, 32 people were approached using various professional networks both online and offline. They represent a broad selection of executives and senior managers across a range of industries. A total of 13 agreed to participate; Table 4 presents their demographics.

| Ref. | Role | Years' experience | Industry | Organisation Size (Turnover) |
|------|------|-------------------|----------|------------------------------|
| P1 | Chief Risk and Compliance Officer | 30+ in Security/IT | Financial Services | >£1B |
| P2 | Chief Information Officer | 30+ in Security/IT | Retail | >£1B |
| P3 | Information Security Manager | 6+ in Security/IT | Logistics | >£1B |
| P4 | Head of Cyber Security | 30+ in Security/IT | Retail | >£1B |
| P5 | Head of IT Operations and Security | 20+ in Security/IT | Manufacturing | >£1B |
| P6 | Head of Information Security | 13+ in Security/IT | Financial Services | >£1B |
| P7 | Director | 20+ in Security/IT | Professional Services | <£1M |
| P8 | Director | 30+ in Security/IT | Professional Services | >£1B |

| P9 | Chief Information Security Officer | 20+ in Security/IT | Professional Services | £500M-£1B |
| P10 | Chief Information Security Officer | 15+ in Security/IT | Not for Profit | >£1B |
| P11 | Director | 30+ in Security/IT | Professional Services | <£1M |
| P12 | Director | 30+ in Corporate Communications / Public Relations | Professional Services | <£1M |
| P13 | Head of Crisis Management | 27+ in Corporate Communications / Public Relations | Professional Services | £500M-£1B |

Table 4 – Demographic information of interview participants

To allow for a more meaningful discussion, prior to the interview, each participant received an interview pack detailing the aims of the framework and the framework itself. The questions asked were based on the framework, its potential value and applicability in supporting businesses in corporate communications in the event of a cyber incident. The interviews lasted up to one hour, with each recorded and transcribed, before then being analysed using thematic analysis. This study received ethical approval through our institution's institutional review board (IRB).

## 7.2 Findings and Discussion

*7.2.1 Framework Feedback*

From an analysis of interviewee responses, we found that perceptions were overwhelmingly positive. In particular, interviewees complimented the structure and the comprehensive nature of the framework. For instance, when asked about their reaction to the framework P11 commented:

> *"I think it is very well-conceived. Very well structured, and I think should absolutely result in good decisions around communications being made."*

Two participants (P7, P13) independently articulated that the framework crystallised their thinking. This suggests that the framework resonated with these professionals and also that they ascribed to the practices it put forward. In particular, several participants felt the framework would help them improve their cyber crisis response by providing them with a codified structure to facilitate the rollout of crisis communication capability. Other interviewees felt that they did not need to change their approach to crisis communications following a data breach in light of this research, as the findings aligned with their current practice. P4, for instance, stated:

> *"We might not have it written down like this or in this type of model. But I think we probably mirror this quite a lot in how we actually practice."*

This is an interesting reaction which, upon follow-up, highlighted a clear endorsement of the framework generally.

The few interviewees who said they would change their current practice (in light of the framework) admitted this because they wanted to adopt what they viewed as the best practice which it aptly captured. Others felt they should adopt the framework within their organisations or for their customers (e.g., in the case of public relations service firms). In some instances, this appeared to be as a result of a perceived lack of maturity in their current approach. On these bases, therefore, it can be seen that the framework was welcomed by these professionals as a novel and informative approach to outlining security incident communication practice. Given the level of industry experience present in the participants interviewed, it is also encouraging that many were interested in adopting the approach and all but one made unsolicited requests to be provided with the final version of the framework.

Of particular interest were the aspects of the framework that interviewees took time to explore in greater detail. For instance, speaking about the framework's pre-event stage, some participants viewed the framework's guidance as crucial as they allowed organisations to think through how they would communicate given different scenarios, agree on the wording of communication and rehearse incident response. P4 commented:

> *"Rehearse it; we talk about muscle memory… Once you've been through that [rehearsals and practicing corporate response] with the right people you don't have to spend a lot of time on it again. We know the answer's "no", because we remember we went through this."*

The importance and implications of the multijurisdictional nature of breaches was another aspect of the framework's stages welcomed by participants. P6 and P7 discussed their own experiences of the various regulatory regimes highlighting the contrasting approaches of different countries, whilst others such as P13 were more unconcerned as they felt they would utilise specialist legal advice if required. They did not, however, disagree with the need to consider jurisdictional issues as part of their crisis response. This point, as well as the others highlighted above, provide an encouraging review of the framework, by specialists with substantial experience and expertise.

It was notable that participants were in support of organisations taking responsibility for breaches and not using SCCT strategies, i.e., framing the organisation as purely being the victim. P11 noted:

> *"When I read it [SCCT], I completely disagreed with it. And it's never a practice that we have put in place for our clients. I guess what I'm saying is that I completely concur with your conclusion."*

Indeed, a majority of interviewees were adamant that organisations were ill-advised to portray themselves as victims during a cyber crisis. This was reinforced by the discourse on responsibility where an overwhelming majority of the participants supported the view in the framework that organisations should accept responsibility for the data loss. Some of the complexities prevalent within supply chains – where data was lost by a partner – were pointed out by participants, but the predominant view was that public-facing organisations should accept responsibility as customers entrusted them to be custodians of their data.

Many participants agreed with the need to front communications via the CEO. P2 elaborated on this highlighting the importance of ensuring they are briefed with the right information to be able to deliver the message agreed by the board. One participant, P6, added the caveat that the CEO needed to be capable of delivering the message effectively, whilst P11 discussed the need to ensure the CEO is only involved in cyber crises (instead of smaller data breaches) as otherwise, this may exaggerate the seriousness of a low impact incident. The framework also seemed to align with the methods used by interviewees to communicate, with P12 recalling use of both social media and email to contact affected parties. The use of telephone communication was also covered by participants with both P9 and P12 highlighting contacting people by telephone as a viable medium.

A final notable point highlighted by P1 was that the framework also considers communication with internal staff in cases of cyber crises:

> *"So, one of the things I was pleased to see was about briefing internal staff. It's actually in my experience one that kind of gets forgotten, is when companies and boards, in particular, spend a lot of time worrying about reputational damage, stock market values, partner things and forget that their employees can feel, why didn't anybody tell me"*.

This is a salient finding for our work as it demonstrates that the framework addresses a current challenge in practice, and could potentially add real value to the industry.

*7.2.2 Refining the framework*

Although the feedback on the framework was positive, participants also provided critiques for aspects of the framework. These were grouped into themes and were reviewed with the aim of improving the framework. This review considered factors such as the number of participants aligned to the theme and the prominence of the theme within the transcripts. Where there were obvious omissions, or where reasonable amendments were warranted, these were also used to enhance the framework. In total, 18 amendments were made. A selection representing themes of particular interest are discussed in the remainder of this section, before then presenting the refined framework.

A significant number of interviewees focused on the security basics guidance within the initial framework. P8, for instance, felt that organisations needed to establish a full range of countermeasures, whereas P7 critiqued each of the measures outlined and P9 felt two-factor authentication was outmoded. P12, however, suggested that the framework should build on existing security standards thereby incorporating best practice. Including a section on security basics in the initial framework was intended to induce organisations to consider obvious security controls pre-event to ensure they had a compelling narrative should a breach occur. Although these comments may be due to the background of interviewees, the security basics guidance seemed to represent a distraction from the communication focus of the framework. This pre-event guidance was therefore reframed to focus on assessing security gaps thus avoiding debate on security basics whilst still prompting organisations to ensure any security weaknesses are reflected in their cyber crisis communication preparations and messaging.

Another area was testing and rehearsals. Both P2 and P4 were strong advocates of rehearsals as a key element of crisis planning and preparation. They highlighted its importance in improving decision making and facing up to the difficult judgement calls associated with data breach crisis management. In particular, the involvement of the CEO in rehearsals and the use of simulations for honing media communications were advocated as good communications practice.

> *"We didn't quite do a media interview but we had a media guy in the simulation who would ask questions like this and our CEO would say, and 'I would answer questions like this'. It was a bit third-party. They sort of rehearsed." [P4]*

Other participants also discussed rehearsals and testing in the context of their current practice. P10, for instance, was concerned that the regular readiness exercises he convened were not taken seriously by some senior stakeholders. P11 felt that organisations that had properly trained for cyber crises were more likely to 'perform well'. These references to rehearsals, training and testing by different interviewees appear to ascribe to the importance of this aspect of pre-event preparation. Although involving suppliers in rehearsals were included in the draft framework on reflection this did not give it adequate coverage and it was therefore given its own section in the updated framework.

There was also another key point related to supporting the board. Providing the CEO with an opportunity to test out media strategies aligns with points made by P1. Like a number of the other participants, P1 felt that the communication should come from the top of the organisation, but he also suggested that the message would be developed through deliberation by the board of directors. To be able to do this effectively, the participant argued, the company's officers required support and education, as well as governance. This was needed to provide structure to the way data breach incidents were handled to prevent them being immediately escalated to the top without due consideration, along with help with framing the message and dealing with complexities such as multi-jurisdictional breaches. He was particularly emphatic on the importance of recognising when, as a group, they were out of their depth and so should seek external support.

P2 and P4, on the other hand, indicated that they would draw on internal expertise such as their public relations and legal departments. This difference in approach could be attributable to the sector of their organisations. Another possibility is that these other areas do make use of external consultants in similar situations, but P2 and P4 are not aware of this due to an IT remit. We should also generally note the influence of an organisation's size as larger enterprise are able to maintain specialist functions due to economies of scale. That notwithstanding, this point seems to represent an additional area for consideration for inclusion in the framework.

The difficulties in assessing the size of data breaches were also discussed by some participants. P2 described the variation in the number of impacted customers typical of a data breach escalation.

> *"We always start off, or we seem to start off with a view from the monitoring that 500,000 customers have been hacked and then you let that play out. Then as you further analyse what's going on and you've really got into it, you end up with 50."*

Whereas P13 believed that often the exact scope of a compromise may never be known. These views were supported by P11 who highlighted the scarcity of information available during a cyber crisis and challenged the precept within the framework that it was better to over rather than under-report on the size of a breach. Although this stance was softened following discussion and P11 was the only interviewee to directly raise this point, the advice to over-report, on reflection, appears too strong. A softer position, counselling organisations to avoid underestimating, is therefore used in the revised framework.

Considering this wide range of feedback from interviewees presented above and in our more complete analysis, we reflected on and refined our framework. The updated version is depicted in Figure 6. This revision also incorporates the fact that interviewees seemed distracted by the fact that the framework initially consisted of separate images. We therefore now have combined these into one, coherent framework. This, we believe, will be more accessible to practitioners.

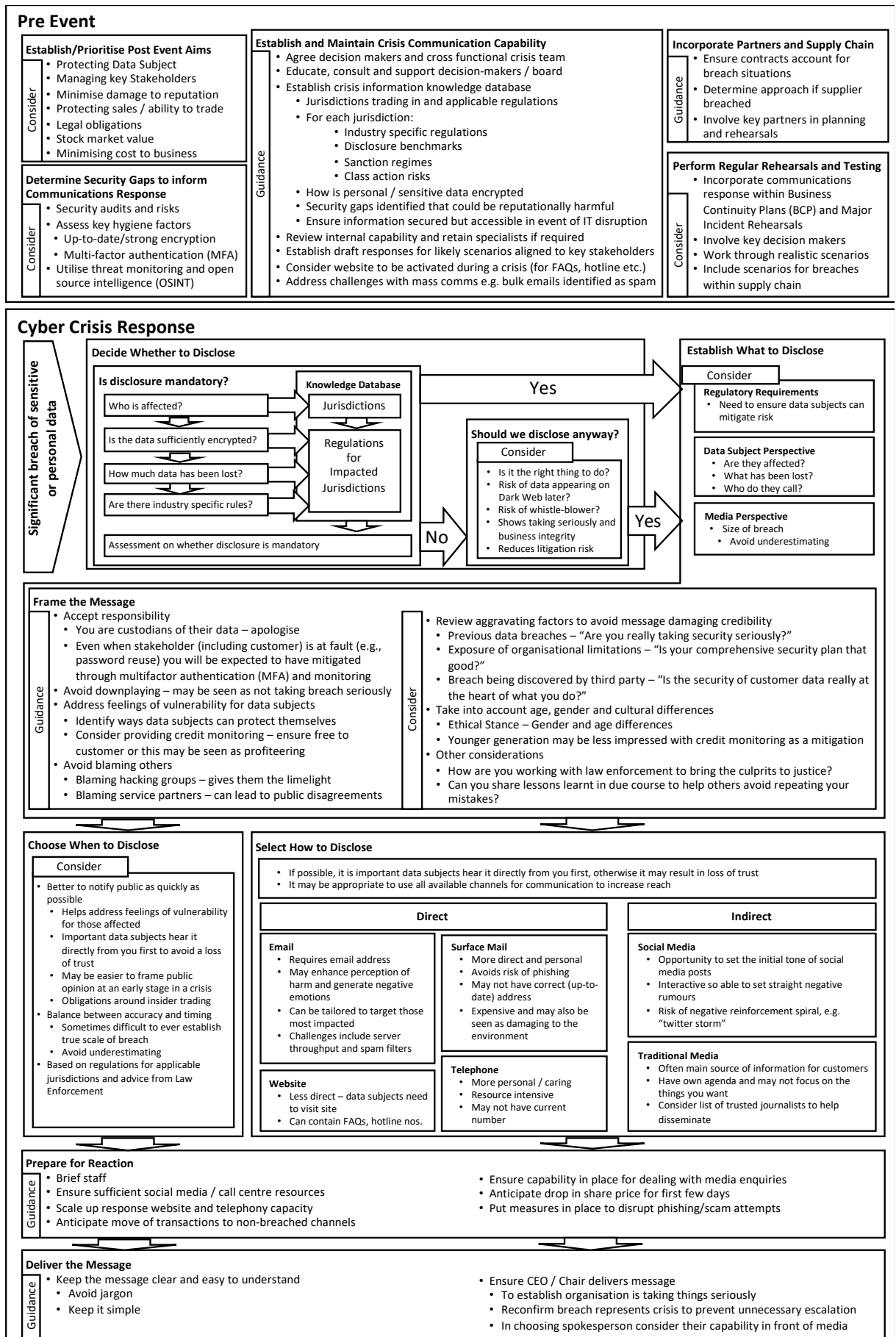

Figure 6. Refined Framework

In general, we can see that many of the previous components are maintained but a number of minor changes have been made to the visual. These are intended to further support organisations and provide a platform for appropriate corporate communications following a significant cyber security data breach. Such a framework could be aligned with wider business continuity approaches and those targeted at security incident management and response within an organisation.

## 8. Conclusion and Future Work

This paper has examined the topical issue of cyber crisis communication, with the intention of improving the understanding of what constitutes effective external communication following data breaches. Furthermore, we have proposed, validated and refined a novel framework which is able to provide guidance to businesses in preparing for and developing appropriate corporate communication responses to such breaches. As such, this work has the potential for significant impact in supporting industry and in providing a basis for more rigorous discussion in academic practice. Another key emerging area where this work may also be useful is in cyber insurance activities associated with breaches. Cyber insurance providers are increasingly relied upon to support businesses after a breach, with services such as public relations and reputation management (Nurse et al., 2020). Our framework could provide a robust and validated structure for such corporate communications plans, which is accessible and can be refined according to each business' context and incident management strategy.

There are some limitations of our work which should be noted. Qualitative data gathering and analysis techniques (e.g., sampling, interviews and thematic analysis), albeit mainstream, can introduce some subjectivity into research studies. We have sought to mitigate this through a rigorous application of these approaches. For instance, in the selection of interviewees, we primarily engaged with senior experts unknown to the researchers to avoid receiving favourable responses due to pre-existing relationships. Moreover, to avoid bias in theme definition, in cases of disagreement assessors discussed and coalesced on appropriate themes. Secondly, this research has been conducted against UK-centric case studies and interviews held with UK-based executives and senior managers. Whilst the systematic literature review references articles produced by authors resident in a number of countries, because of the significant UK emphasis of this work, further investigation would be required to investigate whether this research is directly applicable to data breaches outside of the UK. Work could, however, be undertaken to extend this research to include case studies from other regions, which may provide insight into whether these findings are transferable or any adaptations that need to be made. A third limitation pertains to the use of commentary of a select set of individuals to determine what is effective communication. There may, for instance, be other individuals suitably qualified who hold different valid opinions. We attempted to balance this issue by using well-regarded sources, and our approach of triangulation where we rely on a variety of sources to inform our framework.

For future work we plan to trial, and potentially pilot, the framework within a company with the aim of both integrating it into existing processes and understanding its effectiveness through its application in scenario-based rehearsals. This would provide a crucial practical assessment of its effectiveness and allow examination for any areas of further improvement. We also intend to reflect on the framework and understand the extent to which it is suitable to cater to unintentional data breaches. As mentioned earlier, the framework's guidance does appear to be largely applicable but this will need to be validated, and if necessary, the framework adapted. These efforts can also take advantage of our planned trial and pilot. The framework generally stands to be of great value as it can allow business, cyber security and public relations professionals to plan for and react effectively to major data breaches to avoid their cyber crisis turning into a reputational disaster.

# Author Biographies

**Richard Knight** is a research student in Cyber Security in WMG at the University of Warwick, UK. His research interests include cyber security, corporate communications, security governance, and security in agile development. Richard also has over 20 years of experience in industry spanning various IT and corporate support roles.

**Jason R. C. Nurse** is an Associate Professor in Cyber Security in the School of Computing at the University of Kent, UK. He also holds the role of Visiting Academic at the University of Oxford, UK. He received his PhD from the University of Warwick, UK in 2010. His research interests include security risk management, corporate communications and cyber security, cyber resilience, and insider threat. Jason was selected as a Rising Star for his research into cybersecurity, as a part of the UK's Engineering and Physical Sciences Research Council's Recognising Inspirational Scientists and Engineers (RISE) awards campaign.

# Declaration of Competing Interest

None.